
%
%

\input psfig



%
%
%
%
%


\hsize=6.0truein
\vsize=8.5truein
\voffset=0.25truein
\hoffset=0.1875truein
\tolerance=1000
\hyphenpenalty=500
\def\monthintext{\ifcase\month\or January\or February\or
   March\or April\or May\or June\or July\or August\or
   September\or October\or November\or December\fi}


\font\tenrm=cmr10 scaled \magstep1   \font\tenbf=cmbx10 scaled \magstep1
\font\sevenrm=cmr7 scaled \magstep1  
\font\fiverm=cmr5 scaled \magstep1   

\font\teni=cmmi10 scaled \magstep1   \font\tensy=cmsy10 scaled \magstep1
\font\seveni=cmmi7 scaled \magstep1  \font\sevensy=cmsy7 scaled \magstep1
\font\fivei=cmmi5 scaled \magstep1   \font\fivesy=cmsy5 scaled \magstep1

\font\tentt=cmtt10 scaled \magstep1
\font\tenit=cmti10 scaled \magstep1
\font\tensl=cmsl10 scaled \magstep1

\def\twelvepoint{\def\rm{\fam0\tenrm}
   \textfont0=\tenrm \scriptfont0=\sevenrm \scriptscriptfont0=\fiverm
   \textfont1=\teni  \scriptfont1=\seveni  \scriptscriptfont1=\fivei
   \textfont2=\tensy \scriptfont2=\sevensy \scriptscriptfont2=\fivesy
   \textfont\itfam=\tenit \def\it{\fam\itfam\tenit}
   \textfont\ttfam=\tentt \def\tt{\fam\ttfam\tentt}
   \textfont\bffam=\tenbf \def\bf{\fam\bffam\tenbf}
   \textfont\slfam=\tensl \def\sl{\fam\slfam\tensl} \rm
   \hfuzz=1pt\vfuzz=1pt
   \setbox\strutbox=\hbox{\vrule height 10.2pt depth 4.2pt width 0pt}
   \parindent=24pt\parskip=1.2pt plus 1.2pt
   \topskip=12pt\maxdepth=4.8pt\jot=3.6pt
   \normalbaselineskip=14.4pt\normallineskip=1.2pt
   \normallineskiplimit=0pt\normalbaselines
   \abovedisplayskip=13pt plus 3.6pt minus 5.8pt
   \belowdisplayskip=13pt plus 3.6pt minus 5.8pt
   \abovedisplayshortskip=-1.4pt plus 3.6pt
   \belowdisplayshortskip=13pt plus 3.6pt minus 3.6pt
   \topskip=12pt \splittopskip=12pt
   \scriptspace=0.6pt\nulldelimiterspace=1.44pt\delimitershortfall=6pt
   \thinmuskip=3.6mu\medmuskip=3.6mu plus 1.2mu minus 1.2mu
   \thickmuskip=4mu plus 2mu minus 1mu
   \smallskipamount=3.6pt plus 1.2pt minus 1.2pt
   \medskipamount=7.2pt plus 2.4pt minus 2.4pt
   \bigskipamount=14.4pt plus 4.8pt minus 4.8pt}

\twelvepoint



\font\titlerm=cmr10 scaled \magstep3
\font\titlerms=cmr10 scaled \magstep1 
\font\titlei=cmmi10 scaled \magstep3  
\font\titleis=cmmi10 scaled \magstep1 
\font\titlesy=cmsy10 scaled \magstep3 	
\font\titlesys=cmsy10 scaled \magstep1  
\font\titleit=cmti10 scaled \magstep3	
\skewchar\titlei='177 \skewchar\titleis='177 
\skewchar\titlesy='60 \skewchar\titlesys='60 

\def\titlefont{\def\rm{\fam0\titlerm}
   \textfont0=\titlerm \scriptfont0=\titlerms 
   \textfont1=\titlei  \scriptfont1=\titleis  
   \textfont2=\titlesy \scriptfont2=\titlesys 
   \textfont\itfam=\titleit \def\it{\fam\itfam\titleit} \rm}


\def\preprint#1{\baselineskip=19pt plus 0.2pt minus 0.2pt \pageno=0
   \begingroup
   \nopagenumbers\parindent=0pt\baselineskip=14.4pt\rightline{#1}}
\def\title#1{
   \vskip 0.9in plus 0.45in
   \centerline{\titlefont #1}}
\def\secondtitle#1{}
\def\author#1#2#3{\vskip 0.9in plus 0.45in
   \centerline{{\bf #1}\myfoot{#2}{#3}}\vskip 0.12in plus 0.02in}
\def\secondauthor#1#2#3{}
\def\addressline#1{\centerline{#1}}
\def\abstract{\vskip 0.7in plus 0.35in
	\centerline{\bf Abstract}
	\smallskip}
\def\finishtitlepage#1{\vskip 0.8in plus 0.4in
   \leftline{#1}\supereject\endgroup}

\def\date#1{\finishtitlepage{#1}}

\def\nolabels{\def\eqnlabel##1{}\def\eqlabel##1{}\def\figlabel##1{}%
	\def\reflabel##1{}}
\def\writelabels{\def\eqnlabel##1{%
	{\escapechar=` \hfill\rlap{\hskip.11in\string##1}}}%
	\def\eqlabel##1{{\escapechar=` \rlap{\hskip.11in\string##1}}}%
	\def\figlabel##1{\noexpand\llap{\string\string\string##1\hskip.66in}}%
	\def\reflabel##1{\noexpand\llap{\string\string\string##1\hskip.37in}}}
\nolabels


\global\newcount\secno \global\secno=0
\global\newcount\meqno \global\meqno=1
\global\newcount\subsecno \global\subsecno=0

\font\secfont=cmbx12 scaled\magstep1

\def\section#1{\global\advance\secno by1
   \xdef\secsym{\the\secno.}
   \global\subsecno=0
   \global\meqno=1\bigbreak\medskip
   \noindent{\secfont\the\secno. #1}\par\nobreak\smallskip\nobreak\noindent}

\def\subsection#1{\global\advance\subsecno by1
\medskip
\noindent
{\bf\the\secno.\the\subsecno\ #1}
\par\medskip\nobreak\noindent}

\def\newsec#1{\global\advance\secno by1
   \xdef\secsym{\the\secno.}
   \global\meqno=1\bigbreak\medskip
   \noindent{\bf\the\secno. #1}\par\nobreak\smallskip\nobreak\noindent}
\xdef\secsym{}

\def\appendix#1#2{\global\meqno=1\xdef\secsym{\hbox{#1.}}\bigbreak\medskip
\noindent{\bf Appendix #1. #2}\par\nobreak\smallskip\nobreak\noindent}

\def\acknowledgements{\bigbreak\medskip\centerline{\bf
   Acknowledgements}\par\nobreak\smallskip\nobreak\noindent}


\def\eqnn#1{\xdef #1{(\secsym\the\meqno)}%
	\global\advance\meqno by1\eqnlabel#1}
\def\eqna#1{\xdef #1##1{\hbox{$(\secsym\the\meqno##1)$}}%
	\global\advance\meqno by1\eqnlabel{#1$\{\}$}}
\def\eqn#1#2{\xdef #1{(\secsym\the\meqno)}\global\advance\meqno by1%
	$$#2\eqno#1\eqlabel#1$$}


\def\myfoot#1#2{{\baselineskip=14.4pt plus 0.3pt\footnote{#1}{#2}}}
\global\newcount\ftno \global\ftno=1
\def\foot#1{{\baselineskip=14.4pt plus 0.3pt\footnote{$^{\the\ftno}$}{#1}}%
	\global\advance\ftno by1}


\global\newcount\refno \global\refno=1
\newwrite\rfile

\def\ref{[\the\refno]\nref}
\def\nref#1{\xdef#1{[\the\refno]}\ifnum\refno=1\immediate
	\openout\rfile=refs.tmp\fi\global\advance\refno by1\chardef\wfile=\rfile
	\immediate\write\rfile{\noexpand\item{#1\ }\reflabel{#1}\pctsign}\findarg}
\def\findarg#1#{\begingroup\obeylines\newlinechar=`\^^M\passarg}
	{\obeylines\gdef\passarg#1{\writeline\relax #1^^M\hbox{}^^M}%
	\gdef\writeline#1^^M{\expandafter\toks0\expandafter{\striprelax #1}%
	\edef\next{\the\toks0}\ifx\next\null\let\next=\endgroup\else\ifx\next\empty%

\else\immediate\write\wfile{\the\toks0}\fi\let\next=\writeline\fi\next\relax}}
	{\catcode`\%=12\xdef\pctsign{
\def\striprelax#1{}

\def\semi{;\hfil\break}
\def\addref#1{\immediate\write\rfile{\noexpand\item{}#1}} 

\def\listrefs{\vfill\eject\immediate\closeout\rfile
   {{\secfont References}}\bigskip{\frenchspacing%
   \catcode`\@=11\escapechar=` %
   \input refs.tmp\vfill\eject}\nonfrenchspacing}

\def\startrefs#1{\immediate\openout\rfile=refs.tmp\refno=#1}


\global\newcount\figno \global\figno=1
\newwrite\ffile
\def\fig{\the\figno\nfig}
\def\nfig#1{\xdef#1{\the\figno}\ifnum\figno=1\immediate
	\openout\ffile=figs.tmp\fi\global\advance\figno by1\chardef\wfile=\ffile
	\immediate\write\ffile{\medskip\noexpand\item{Fig.\ #1:\ }%
	\figlabel{#1}\pctsign}\findarg}

\def\listfigs{\vfill\eject\immediate\closeout\ffile{\parindent48pt
	\baselineskip16.8pt{{\secfont Figure Captions}}\medskip
	\escapechar=` \input figs.tmp\vfill\eject}}

\def\noblackbox{\overfullrule=0pt}
\def\inv{^{\raise.18ex\hbox{${\scriptscriptstyle -}$}\kern-.06em 1}}
\def\dup{^{\vphantom{1}}}
\def\Dsl{\,\raise.18ex\hbox{/}\mkern-16.2mu D} 
\def\dsl{\raise.18ex\hbox{/}\kern-.68em\partial}
\def\slash#1{\raise.18ex\hbox{/}\kern-.68em #1}
\def\lspace{}
\def\lbspace{}
\def\boxeqn#1{\vcenter{\vbox{\hrule\hbox{\vrule\kern3.6pt\vbox{\kern3.6pt
	\hbox{${\displaystyle #1}$}\kern3.6pt}\kern3.6pt\vrule}\hrule}}}
\def\mbox#1#2{\vcenter{\hrule \hbox{\vrule height#2.4in
	\kern#1.2in \vrule} \hrule}}  
\def\bar{\overline}
\def\e#1{{\rm e}^{\textstyle#1}}
\def\del{\partial}
\def\curly#1{{\hbox{{$\cal #1$}}}}
\def\curlyD{\hbox{{$\cal D$}}}
\def\curlyL{\hbox{{$\cal L$}}}
\def\vev#1{\langle #1 \rangle}
\def\psibar{\overline\psi}
\def\lform{\hbox{$\sqcup$}\llap{\hbox{$\sqcap$}}}
\def\darr#1{\raise1.8ex\hbox{$\leftrightarrow$}\mkern-19.8mu #1}
\def\half{{\textstyle{1\over2}}} 
\def\roughly#1{\ \lower1.5ex\hbox{$\sim$}\mkern-22.8mu #1\,}
\def\MSbar{$\bar{{\rm MS}}$}
\hyphenation{di-men-sion di-men-sion-al di-men-sion-al-ly}

\parindent=0pt
\parskip=5pt


\preprint{
\vbox{
\rightline{CERN-TH.7281/94}
\vskip2pt\rightline{SHEP 93/94-23}
}
}
\vskip -1cm

\title{On Truncations of the Exact Renormalization Group}
\vskip -1cm
\author{\bf Tim R. Morris}{}{}
\vskip 0.5cm
\addressline{\it CERN TH-Division}
\addressline{\it CH-1211 Geneva 23}
\addressline{\it Switzerland\myfoot{$^*$}{\rm On leave from Southampton
University, U.K.}}
\addressline{\it }
\vskip -1cm

\abstract 
We investigate the Exact Renormalization Group (ERG) description of
($Z_2$ invariant) one-component scalar field theory, in the
approximation in which all momentum dependence is discarded in the
effective vertices.  In this context we show how one can perform a
systematic search for non-perturbative continuum limits without making
any assumption about the form of the lagrangian. The approximation is
seen to be a good one, both qualitatively and quantitatively.
We then consider the {\sl further} approximation of truncating the
lagrangian to polynomial in the field dependence.
Concentrating on
the non-perturbative three dimensional Wilson fixed point, we show
that the sequence of truncations $n=2,3,\dots$, obtained by expanding
about the field $\varphi=0$ and discarding all powers $\varphi^{2n+2}$
and higher, yields solutions that at first converge to the answer
obtained without truncation, but then cease to further converge beyond
a certain point. Within the sequence of truncations,
no completely reliable method exists to reject the
many spurious solutions that are also generated. These properties are
explained in terms of the analytic behaviour of the untruncated
solutions -- which we describe in some detail.
\vskip -1cm
\date{\vbox{
{CERN-TH.7281/94}
\vskip2pt{SHEP 93/94-23}
\vskip2pt{hepth/9405190}
\vskip2pt{May, 1994.}
}
}
\catcode`@=11 
\def\slash#1{\mathord{\mathpalette\c@ncel#1}}
 \def\c@ncel#1#2{\ooalign{$\hfil#1\mkern1mu/\hfil$\crcr$#1#2$}}
\def\lsim{\mathrel{\mathpalette\@versim<}}
\def\gsim{\mathrel{\mathpalette\@versim>}}
 \def\@versim#1#2{\lower0.2ex\vbox{\baselineskip\z@skip\lineskip\z@skip
       \lineskiplimit\z@\ialign{$\m@th#1\hfil##$\crcr#2\crcr\sim\crcr}}}
\catcode`@=12 
\def\nonp{non-perturbative}
\def\phi{\varphi}
\def\te#1{\theta_\epsilon( #1,\Lambda)}
\def\th{\vartheta}
\def\epsilon{\varepsilon}
\def\p{{\bf p}}
\def\P{{\bf P}}
\def\q{{\bf q}}
\def\r{{\bf r}}
\def\x{{\bf x}}
\def\y{{\bf y}}
\def\tr{{\rm tr}}
\def\D{{\cal D}}
\def\iprop{\Delta_\Lambda^{-1}\! }
\def\ins#1#2#3{\hskip #1cm \hbox{#3}\hskip #2cm}
\def\frac#1#2{{#1\over#2}}
\nref\Wil{K. Wilson and J. Kogut, Phys. Rep. 12C (1974) 75.}
\nref\weg{F.J. Wegner and A. Houghton, Phys. Rev. A8 (1973) 401.}
\nref\deriv{T.R. Morris, Phys. Lett. B329 (1994) 241.}
\nref\erg{T.R. Morris, Int. J. Mod. Phys. A9 (1994) 2411.}
\nref\truncm{``Momentum Scale Expansion of Sharp Cutoff Flow
Equations'', T.R. Morris, CERN / Southampton preprint in preparation. }
\nref\trunci{A. Margaritis,
G. \'Odor and A. Patk\'os, Z. Phys. C39 (1988) 109.}
\nref\truncii{P.E. Haagensen, Y.
Kubyshin, J.I. Latorre and  E. Moreno, Phys. Lett. B323 (1994) 330.}
\nref\al{M. Alford, Cornell preprint CLNS 94/1279.}
\nref\wet{N. Tetradis, C.  Wetterich, preprint DESY-93-094.}
\nref\hashas{ A. Hasenfratz and P. Hasenfratz, Nucl. Phys. B270
(1986) 685.}
\nref\others{
U. Ellwanger (et al), Heidelberg preprints HD-THEP- 94-1, 94-2, 93-30, 92-49;\
T.E. Clark et al, Purdue preprints PURD-TH- 94-01, 92-9;\
M. Alford and J. March-Russell, Cornell preprint CLNS 93/1244;\
G. Felder, Commun. Math. Phys. 111 (1987) 101;\
M. Reuter and C. Wetterich, HD-THEP-93-41 and refs therein;\
M. Maggiore, Z. Phys. C41 (1989) 687;\
S-B Liao and J. Polonyi, Duke preprint Duke-TH-94-64.
}
\def\lastref{\others}


In ref.\deriv\ we claimed that a sequence of truncations of the field
dependence of the ERG\Wil\weg\ do not work in
general (in the sense of providing dependable and in principle
arbitrarily accurate results), because these do not converge beyond a
certain point, and because no completely reliable method exists
to reject the many spurious solutions that are also generated.
In this letter 
 we verify this claim for the relatively simple
$O(p^0)$ case described in the abstract. We compute the truncations to
high order ($n=25$), and show how this behaviour may be
understood, accurately and at a deeper level,
 in terms of the analytic behaviour of the untruncated solutions
-- which is described here in much more detail than was possible in ref.\deriv.
(We must emphasise here the distinction between a {\sl momentum/derivative
expansion} in which higher space-time derivative terms are discarded but
{\sl no approximation} is made in the field dependence -- these approximations
{\sl do} appear to converge\erg\deriv\ -- and {\sl truncations} of the field
dependence of the lagrangian which in general do not converge and can give even
qualitatively wrong results).

We point out that the analytic properties of the untruncated solutions
allow for the possibility of searching, within the momentum/derivative
expansion \deriv--\truncm, {\sl systematically}
through the infinite dimensional space of non-perturbative lagrangians
for new continuum limits.
It need hardly be stated that, firstly very little is known about the
possible existence of continuum theories in such a space, and secondly,
if new fixed points were  found, they could have profound implications.
We have done
such a search for $O(N)$ scalar field theory in $D=$ 4 and 3
dimensions, for the cases $N=1,2,3,4$, in the $O(p^0)$
approximation.  However, we found there to be
only the known fixed points at this level:
Gaussian for $D=3$ and 4, and the Wilson fixed points in 3 dimensions.

Finally we construct two better methods of approximation by expansion.
 The simplest way to
solve the $O(p^0)$ equations, is  directly
numerically\deriv\ however (see also later, and ref.\hashas).
The truncations, if carefully interpreted, can give moderately accurate
results -- thus the simpler low order truncations may be of some use
 in situations where more reliable and
accurate calculations are prohibitively difficult to perform.
This situation seems very reminiscent of approximations (also
involving truncations of the operator basis) to the ``real space
renormalization group'' investigated in the late 1970's\ref\realsp{See
for example the reviews in  eds. T.W. Burkhardt and J.M.J. van Leeuwen,
``Real-Space Renormalization'' (1982), Springer, Berlin. }.
For  recent work on approximations
to the ERG see for example refs.\deriv--\lastref; From Alford's work\al,
and the numerical solution, it is
 clear that expansion of $\phi$ around the semi-classical minimum of
the effective potential, results at low orders of truncation,
in convergence to three decimal places or more for the $O(p^0)$ $\nu$
(or the other exponents related by scaling).
It would be interesting to better understand
the reliability of this method and its limiting
accuracy (also for $\omega$). Its behaviour no doubt is otherwise
similar to the truncations we discuss, and for the same
underlying causes. Effectively this expansion was part of the
calculation in ref.\wet, where the authors claim to
obtain quite
accurate values for exponents of $3D$ $N$-vector models. At higher orders
in the momentum expansion we expect that truncations become more
limited in accuracy and reliability,
if only because it ought now to involve the much more complicated
calculation of the expansion of several functions\deriv\ in
polynomials, each with their own limitations in accuracy. Indeed,
 even in ref.\truncm, where we will consider truncations in the field
dependence of higher momentum terms ($O(p^n)$ with $n>0$) only, making
no expansion in the potential, the results support this hypothesis.

Here we will concentrate on the case of sharp cutoff and $O(p^0)$, i.e.
the remarkable equation\ref\orig{J.F. Nicoll, T.S. Chang and H.E.
Stanley, Phys. Lett. 57A (1976) 7.}\ first studied without further
approximation, by Hasenfratz and Hasenfratz\hashas. We start with a
simple alternative derivation.\foot{ For more information on this, see
ref\erg.}\ The equivalent smooth cutoff equation\deriv\ is
qualitatively very similar so that much the same analysis,
and all our general conclusions, apply
equally well to the smooth case. We work in $D$
euclidean dimensions with a single real scalar field $\phi$.  The
partition function is defined as
\eqn\zorig{\exp W_\Lambda[J]=\int\!\D\phi\
\exp\{-\half\phi.\iprop.\phi-S_{\Lambda_0}[\phi]+J.\phi\}\ \ .}
 $\Delta_\Lambda\equiv\Delta(q,\Lambda)=\te q/q^2$ is a
free massless propagator times a smooth (everywhere positive)
 infrared cutoff function $\te q$. The sharp cutoff limit is given by
the Heaviside function:
\eqn\thelimit{\te q\to \theta(q-\Lambda)\ins11{as} \epsilon\to0\ \ .}
{}From \zorig\ we derive the flow equation for $W_\Lambda$:
$${\partial\over\partial\Lambda}W_\Lambda[J]=
-{1\over2}\left\{ {\delta W_\Lambda\over
\delta J}.{\partial \Delta_\Lambda^{-1} \over\partial\Lambda}.{\delta
W_\Lambda\over\delta J} +
\tr\left({\partial \Delta_\Lambda^{-1} \over\partial\Lambda}.{\delta^2
W_\Lambda\over\delta J\delta J}\right)\right\}\quad ,$$
which on rewriting in terms of the interaction part of the Legendre effective
action via $\Gamma_\Lambda[\phi]+\half\phi.\iprop.\phi
=-W_\Lambda[J]+J.\phi$,\quad $\phi=\delta W_\Lambda/\delta J$, gives
\eqn\fgam{ {\partial\over\partial\Lambda}\Gamma_\Lambda[\phi]=
-{1\over2}\tr\left[{1\over \Delta_\Lambda}{\partial
\Delta_\Lambda\over \partial\Lambda}
.\left( 1+\Delta_\Lambda.{\delta^2\Gamma_\Lambda\over\delta\phi\delta\phi}
\right)^{-1}\right]\ \ .}
$\Gamma_\Lambda$ generates the one particle irreducible parts of the Wilson
effective action (Wegner-Houghton effective action\weg\ in the sharp cutoff
limit) \erg.
If we now make the approximation of discarding all momentum
dependence from $\Gamma_\Lambda$ in \fgam, we obtain
\eqn\inter{{\partial\over\partial\Lambda}V(\phi,\Lambda)=-{1\over2}
\int\!{d^Dq\over(2\pi)^D}\,\left\{ {\partial\te q\over\partial\Lambda}
{1\over \te q \left[ 1+\te q V''(\phi,\Lambda)/q^2\right]}\right\}\ \ ,}
where we have introduced the potential through
$\Gamma_\Lambda =\int\!d^Dx\,
V(\phi(\x),\Lambda)$. Primes denote differentiation with respect to $\phi$.
The integrand (in curly brackets) in \inter\ may be written
$${\partial\over\partial\Lambda}\left\{ \ln\te q-\ln\left[1+\te q
V''(\phi,{\tilde\Lambda})/q^2\right]
\right\}\Bigm|_{{\tilde\Lambda}=\Lambda}\ \ .$$
The first term above yields a field independent vacuum energy which we drop
from $V$. Taking the sharp cutoff limit \thelimit\ we thus obtain for the
integrand
$$\delta(q-\Lambda)\ln(1+V''(\phi,\Lambda)/q^2)\ \ .$$
The integral in \inter\ is now trivial. Factoring out the scale $\Lambda$
by writing $\phi\mapsto\Lambda^{D/2-1}\phi/\zeta$,
$V(\zeta^{-1}\phi\Lambda^{D/2-1},\Lambda)
\mapsto \zeta^{-2}\Lambda^DV(\phi,t)$, and  $t=\ln(\Lambda_0/\Lambda)$, where
the factor $\zeta=(4\pi)^{D/4}\sqrt{\Gamma(D/2)}$ is chosen for convenience,
we obtain the advertised equation\orig\hashas:
\eqn\Veq{{\partial\over\partial t}V(\phi,t)+(D/2-1)\phi
V'(\phi,t)-D V(\phi,t)= \ln\left[1+V''(\phi,t)\right]\ \ .}
 The anomalous dimension $\eta=0$
in this case since a non-zero $\eta$ results from non-trivial
wavefunction renormalization.  The reader may be puzzled as to why
we obtain the same equation as that for the Wilson effective potential\hashas\
 given that ours is the Legendre
effective potential. In fact these are one and the same, since at zero momentum
all vertices of the Wilson effective action $S_\Lambda$
coincide with $\Gamma_\Lambda$ \erg.


Now we set $D=3$. From \Veq, a fixed point
effective potential $V(\phi,t)\equiv V(\phi)$ must satisfy
\eqn\fp{\eqalign{
{1\over2}\phi V'(\phi)-3 V(\phi) &= \ln\left[1+V''(\phi)\right]\ \ ,
\hskip 1cm\cr
{\rm and}\hskip 1cm  V'(0)&=0}}
 (by $\phi\leftrightarrow-\phi$ invariance).
At first sight these equations appear to have many solutions, which may be
parametrized by $V(0)=-\ln(1+\sigma)/3$, $\sigma\Lambda^2$ being the
semi-classical effective mass-squared.
Actually this is not the case\hashas,
because {\sl all but two solutions end at a singularity of
the form} $V(\phi)\sim 2(1-\phi/\phi_c)\ln(\phi_c-\phi)$, $\phi_c$ some
positive constant, or more precisely, 
as a series in  decreasingly singular terms,
\eqn\sing{\eqalign{V=&\ln(x)\left(x-{3\over8}x^2-{25\over432}x^3
-{5\phi_c^2\over384}x^4+{3169\over27648}x^4+\cdots\right)\cr
&+\ln(x)^2\left({25\over288}x^3-{25\over1152}x^4-\cdots\right)
+O(\ln(x)^4 x^5)\ \ ,}}
where $x=1-\phi^2/\phi_c^2$. We now justify this statement, by dividing
the behaviour of $V(\phi)$ into three classes.

First of all, if
$V$ ends at a singularity, it, or some derivative of it diverges there. By
considering how
the various terms can balance in \fp, one sees that a singularity
must be of the form above.
Secondly, if $V$ does not end at a singularity but instead it and its
first two derivatives tend to a limit ($\pm\infty$ included) as
$\phi\to \infty$, then again, considering the balance of terms in
\fp\ one sees that either $V(\phi)\to 0$ or $V(\phi)$ satisfies
\eqn\asy{V(\phi)=A\phi^6-{4\over3}\ln\phi -{2\over9} -{1\over3}\ln(30A)
-{1\over150A\phi^4}+O(1/\phi^6)\ \ ,}
for some positive constant $A$,
as $\phi\to\infty$. Linearizing in \fp\
about the first possibility, i.e. setting
$V(\phi)\mapsto V(\phi) +\delta V(\phi)$, one finds that $\delta V$ must behave
for large $\phi$ as a linear combination of $\phi^6$ and $\exp(\phi^2/4)$.
Since both of these corrections are excluded
if we require also $V(\phi)+\delta V(\phi)\to0$ as $\phi\to
\infty$, we conclude that $V(\phi)$ is identically zero in this case,
which indeed is
the trivial Gaussian solution to \fp. (For a study of perturbations about
the Gaussian  in eqn.\Veq, see ref.\hashas). Linearizing about the
second possibility, i.e. taking $V(\phi)$ to be as in \asy, one finds that
$\delta V$ must behave for large $\phi$ as a linear combination of $\phi^6$
and $\exp(5A\phi^6/2)$.  Now the first correction merely perturbs the
coefficient $A$, while the second is excluded if $V+\delta V$ also satisfies
\asy. It follows that the space of
solutions obeying \asy\ divide into {\sl isolated} one-parameter subsets,
each parametrized by $A$. For both possibilities the
exponentially growing corrections were the result of linearizing the singular
behaviour \sing, since they involve balancing the same terms in \fp.
Thirdly, $V$ or one of its first two derivatives may be defined for all
finite real $\phi$ but not tend to a limit as $\phi\to\infty$. Studying
\fp\ one sees that this requires solutions to become infinitely
oscillatory as $\phi\to\infty$. We do not supply a proof that this does
not happen, but we feel confident we can rule it out because we saw no hint
of it in our extensive numerical and analytic investigations (see below).
\midinsert
\centerline{
\psfig{figure=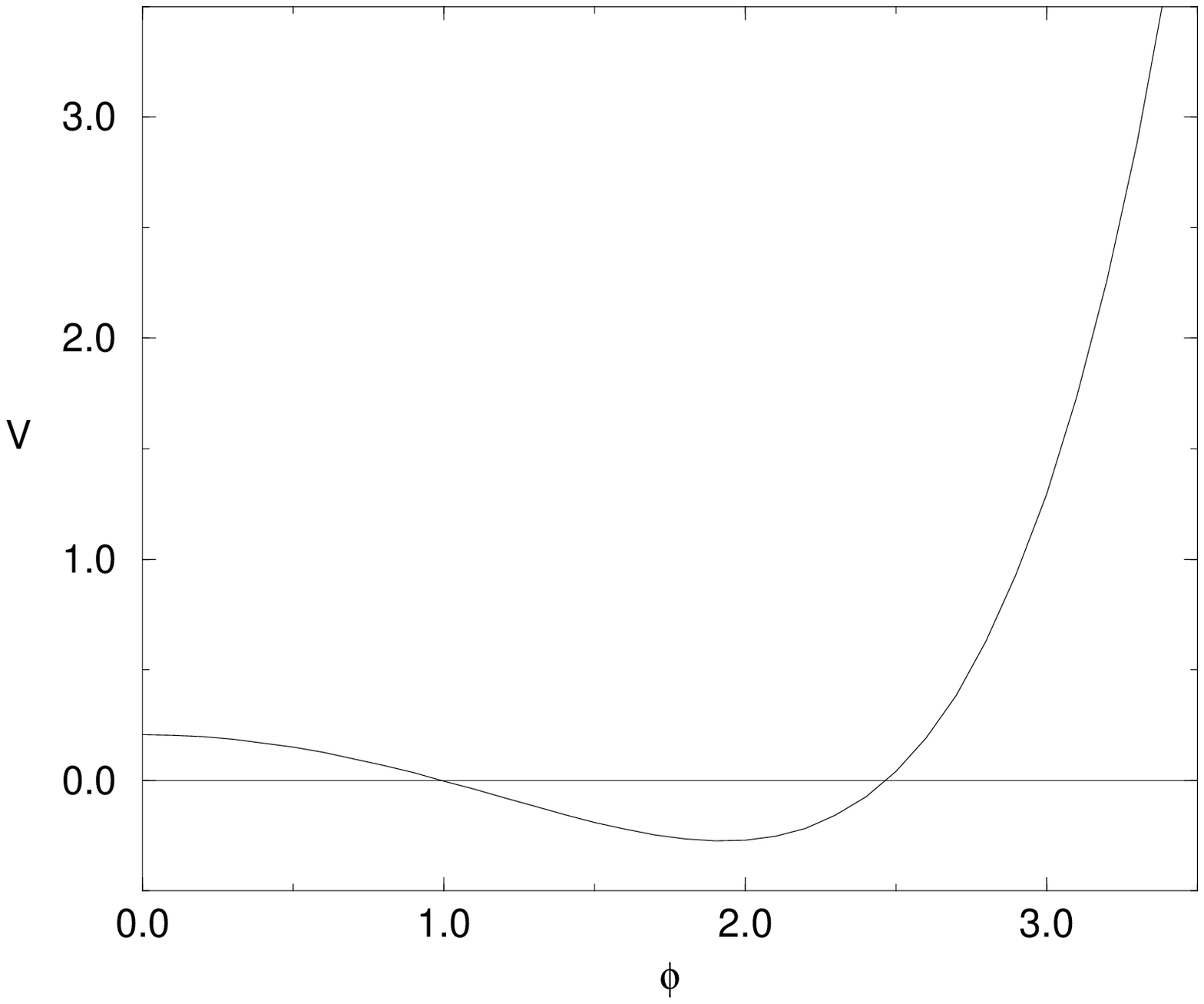,width=4.3in}}  
\vskip 0in
\centerline{\vbox{{\bf Fig.1.} The solution $V(\phi)$ which approximates the
Wilson fixed point.
}}
\endinsert

Therefore, apart from the trivial solution $V(\phi)\equiv0$, any global
fixed point solution of \fp\ must satisfy {\sl two} boundary
conditions: \asy\ and $V'(0)=0$. We thus expect at most a countable
number of such solutions; we find only
one. It may be characterized by $\sigma=\sigma_*=-.46153372\cdots$
(or $A=A_*=.003033\cdots$) and is displayed in fig.\fig\fpV{ }. (We
describe the search 
below).
This is the same solution as in ref.\hashas\ of
course and is a fair approximation to the Wilson fixed point.
To obtain the critical exponents and operator
spectrum one linearizes about this solution in \Veq, i.e. one writes
$V(\phi,t)=V(\phi)+\delta V(\phi,t)$, with $\delta V(\phi,t)\propto
v(\phi) \exp(\lambda t)$. If one excludes the exponentially growing
solution (which is $\exp(5A\phi^6/2)$ again) one obtains a discrete
spectrum, and deduces from the eigenvalues the critical exponents,
in reasonable agreement with the best estimates,
as already described in ref.\deriv\
(cf. also \hashas\ and later). It is surely the case that the
exponentially growing solution is again a linearization of singular
behaviour, this time in the flowing solution $V(\phi,t)$.

Note that it is wrong to conclude from fig.\fpV, e.g. by na\"\i ve
semi-classical considerations, that the Wilson fixed point
describes a spontaneously broken theory: indeed this is inconsistent with
the fact that the field theory is scale invariant at this point. The
remaining quantum corrections for momenta $q<\Lambda$, which generically
give positive contributions to the mass-squared, in this case exactly cancel
the negative mass-squared $\sigma_* \Lambda^2$, and as $\Lambda\to 0$
one  recovers the complete Legendre effective potential as
$V(\phi)=A_*\phi^6$ (from \asy\ in unscaled units).
Note also that it is not necessary to consider separately the question
of the physical stability of solutions to \fp. This is
because there are no solutions that are unbounded from below
 (contrary to the statement in ref.\hashas), while
solutions that end on the singularity \sing\ are clearly physically
unacceptable:  the potential does not {\sl
diverge} at $\phi_c$,  it simply ceases to exist -- or is complex thereafter.
(In the Gaussian case the stability is seen to hold automatically once
perturbations about the fixed point are considered, as a consequence of
the ln in \Veq).

The general structure of the solutions is as follows. For $\sigma$ close but
less than $\sigma_*$,
the solutions look very similar to fig.\fpV\ but end at some $\phi=\phi_c$
in a singularity \sing, as is most easily seen by plotting
$V''\sim 2/\phi_c/(\phi_c-\phi)$. As $\sigma$ approaches $\sigma_*$ from
below, $\phi_c$ moves out along the real axis, but very slowly, so that
for $\phi_c>3$ we require $\sigma_*-\sigma<\Delta$ where $\Delta\approx.005$,
while $\sigma$ must approximate $\sigma_*$ to high precision for say
$\phi_c\gsim4$. For $\sigma$ at the same proximity but above
$\sigma_*$ the singularity splits into a complex conjugate pair close to
the real axis, with real positions Re$(\phi_c)$ in approximately
 the same place. These move closer to, and out along, the real axis as
$\sigma\to\sigma_*^+$. The distance from the real axis is also a sensitive
function of $\sigma-\sigma_*$.
(Perhaps the position of these singularities is given by
$\phi_c\sim (\sigma_*-\sigma)^{-\tau}$, when $\sigma\approx\sigma_*$,
$\sigma{<\atop>}\sigma_*$, for some small positive
constant exponent $\tau$). On the real axis, at $\phi\approx{\rm Re}(\phi_c)$,
$V''$ turns over steeply and drops
rapidly to a value just greater than $-1$ and approximately constant over
a large range. In this range $V$ slowly ``rolls over'' as
\eqn\slorol{V(\phi)=-{1\over3}\ln\delta +c\phi-{1\over2}\phi^2
+\delta\int^\phi_0 \!d\psi\, (\phi-\psi)\exp\{\psi^2-{5\over2}c
\psi\}\quad+\cdots}
where $\phi=c+O(\delta)$
is the point inside the range where the potential reaches a maximum, $\delta$
is small ($\delta\to0$ as $\sigma\to\sigma_*^+$), and
$|\phi-c|<\!\!\!~<\sqrt{\ln(1/\delta)}$.
 Eventually, at some
position $\phi=\phi_c'$ outside this range, $V$ encounters another singularity
of the form \sing.

For $\sigma=-1+\delta$, $\delta\to0^+$, the solution
$V(\phi)$ is governed by \slorol\ with $c=0$, ending in a singularity
$\phi'_c\sim\sqrt{\ln(1/\delta)}$.
As $\sigma\to0^-$, the singularities move out to infinity
in such a way that $V(\phi)$ tends pointwise to zero. For $\sigma>0$,
$V(\phi)$ grows monotonically (with real increasing $\phi$)
ending in a singularity \sing, which moves out slowly
as $\sigma$ is increased.

Returning to the true solution, $V(\phi)$ at $\sigma=\sigma_*$, we note that
$V$ has a four-fold symmetry in the complex plane:
complex conjugation $\times$ ($\phi\leftrightarrow-\phi$).
Factoring out this symmetry, if one
carefully integrates out along rays $\phi=r\,\e{i\th}$, with $0\le \th
\le \pi/2$, one can determine the position of the closest
singularity. It appears at $r=r_*=3.12$, $\th=\th_*=.257\pi$. (There are
others with $r>r_*$).


We see that it is possible to make a systematic search for new continuum
limits {\sl without making any assumption about the form of the bare
potential}.
Indeed if such a potential can be tuned to a critical point where a continuum
limit is recovered, then the corresponding Wilson effective potential must
satisfy \fp\ at that point. On the the other hand, if we find an acceptable
solution of \fp, then because such a solution is scale invariant, this
equally well serves as the critical bare potential. Thus a search over the
infinite dimensional space of bare potentials reduces to a one dimensional
search for effective potentials obeying \fp\ that do not end in a singularity.
In  fig.\fig\searchfp{ }\ we plot the inverse of the
position of the real singularity against $\sigma$.  The Gaussian and
Wilson fixed points are clearly seen as sharp downward spikes, at
$\sigma=0$ and $\sigma=\sigma_*$ respectively. We have performed the
equivalent search in $O(N)$ scalar field theory in $D=3,4$ dimensions for
$N$ from 1 to 4, as mentioned in the introduction -- finding only the
expected fixed points. (The relevant equation for general $N$ is given for
example in ref.\hashas). Of course the restriction to considering only general
potentials is a result of our approximation; at higher orders in the
momentum expansion a larger space of lagrangians can be searched.
\midinsert
\centerline{
\psfig{figure=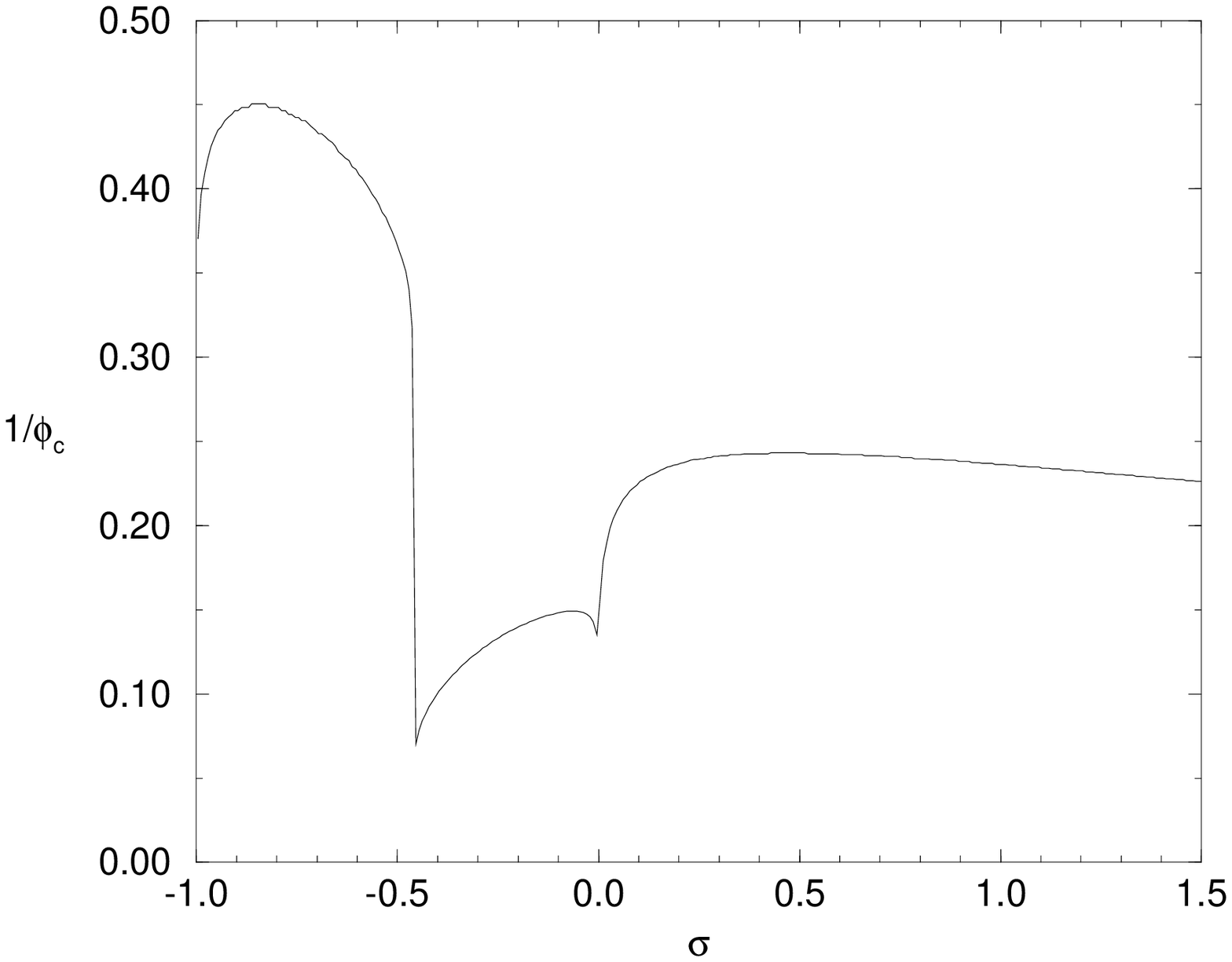,width=4.3in}}  
\vskip 0in
\centerline{\vbox{{\bf Fig.2.} The inverse of the position of the real
singularity $\phi_c$ as a function of $\sigma$, sampled with a stepsize
$\delta\sigma=1/120$. The downward spikes reach further towards zero with
a finer mesh. The various features are explained in the description of the
analytic structure given earlier.
}}
\endinsert

(Note that equation \fp\ is stiff\ref\NR{ For a discussion of stiffness and
what to do about it see e.g. W.H. Press et al, ``Numerical Recipes .. The
Art of Scientific Computing'', 2nd edition (1992) C.U.P.}, the higher
order equations\deriv\ more severely so. To obtain an accurate representation
of the solution at $\sigma=\sigma_*$ one can binary chop between the slow
rollover behaviour \slorol\ for $\sigma>\sigma_*$ and the singular behaviour
\sing\ for $\sigma<\sigma_*$, but a more efficient method is to require
\asy\ at some large value of $\phi$. One can then either shoot to the
origin\foot{For $N\ne1$ one would need to shoot to an intermediate fitting
point.}\ -- determining $A$ to zero $V'(0)$, or use relaxation.)


We can now give an intuitive explanation,
based on the simplified context of \fp, for why the truncations in fact
converge at first but then cease to further
converge beyond a certain maximum $n$.
The point is that, as well as the true solutions,
there are many `bad' solutions with singular field dependence on the real axis.
Very bad solutions have (real or complex)
singular field dependence very close to the origin $\phi=0$,
causing the coefficients of $\phi^m$ ($m$-point Green functions or vertices
in general) to diverge rapidly with $m$ according to the appropriate
radius of convergence. Naturally, the polynomial field dependence of the
truncations, for which the $2n+2$ vertex vanishes,\foot{For a discussion of
truncations in general see for example ref\erg.}\
 tend therefore
to better approximate the Taylor series of a true solution. Increasing $n$
will tend at first to further improve the approximation, by in effect ensuring
that the singularities are forced further from the origin. However a
non-trivial true solution also has singularities for  complex
 $\phi$ at (and in general beyond) some radius $|\phi|=r_*$.
Therefore the truncations cannot be expected to converge to better results
than would be obtained from `moderately bad' solutions with singular field
dependence only at or beyond the radius $r_*$.

Now let us make this argument much more precise.
If we wish to ensure that the potential $V(\phi)$ is an even function, then
the Taylor expansion must be done about the origin. It is helpful to write
\eqn\expa{V(\phi)=-{1\over3}\ln(1+\sigma)+{1\over2}\sigma\phi^2+4\sum_{k=2}
{\alpha_{2k}(\sigma)\over2k(2k-1)}\phi^{2k}\quad.}
Plugging this into eqn.\fp\ we deduce $\alpha_4=-{1\over4}\sigma(1+\sigma)$,
$\alpha_6={1\over48}\sigma(1+\sigma)(1+7\sigma)$,
$\alpha_8=-{1\over48}\sigma^2(1+\sigma)(1+3\sigma)$, $\cdots$, and for $k\ge4$
that $\alpha_{2k}$ is given as $\sigma^2$ times $(1+\sigma)$ times a
polynomial in $\sigma$ -- as follows from $D=3$ being an upper critical
dimension\truncii, the slow rollover behaviour as $\sigma\to-1$, and the
recurrence relation
$${\alpha_{2k+2}\over1+\sigma}={k-3\over2k(2k-1)}\alpha_{2k}
+\sum_{m=2}^k{(-4)\over m}^{m-1}\!\!\!\!\!\!
\sum_{\scriptstyle k_1,\cdots,k_m\ge1
\atop \scriptstyle k_1+\cdots+k_m=k}{\alpha_{2k_1+2}
\cdots\alpha_{2k_m+2}\over(1+\sigma)^m}\quad,$$
respectively.
The $n^{\rm th}$ truncation is defined by setting $\alpha_{2n+2}(\sigma)=0$.
We concentrate on large $n$, where
the solutions $\sigma$ that result from this, can be understood from
the asymptotic expressions for $\alpha_{2k}$. To leading order in $1/k$,
if the closest singularities \sing\ are
on the real axis at $\phi_c=\pm r$
one has $\alpha_{2k}\sim1/r^{2k}$, while if they are complex
then there are four in the form $\phi_c=\pm r\,\exp(\pm i\th)$ and
one has $\alpha_{2k}\sim2\cos(2k\th)/r^{2k}$.
(In fact the $\alpha_{2k}$
asymptote to these expressions even for quite small $k$; for
example one finds for $\sigma=\sigma_*$,
that the
asymptotic expressions give the right sign for $k>2$ and are within
a factor 2 for $k\ge7$). Recalling the analytic behaviour
of $V(\phi)$ as a function of $\sigma\approx\sigma_*$,
we see that for $\sigma<\sigma_*-\Delta$ the
coefficients $\alpha_{2n+2}(\sigma)$
are all positive, so there are no solutions $\sigma$ in this region for
large $n$. On the other hand for $\sigma=\sigma_*$
 the coefficients are controlled by the
singularities $\phi_c=\pm r_*\exp(\pm i\th_*)$ and thus the
signs of the $\alpha_{2n+2}(\sigma_*)$
are very closely four-fold periodic in $n$
in the pattern  $++--$, as a consequence of $\th_*\approx\pi/4$.
For the negative pair, it follows
by continuity in $\sigma$ that $\alpha_{2n+2}(\sigma)$ must
vanish for some $\sigma$ in the range $\sigma_*-\Delta<\sigma<\sigma_*$.
For the positive pair, one has to look at
$\sigma>\sigma_*+\Delta$; here the closest singularities are complex
with an angle $\th$ which is a rapidly increasing function of $\sigma$. This
implies that there is always a zero of $\alpha_{2n+2}(\sigma)$ close to
but greater than $\sigma_*+\Delta$. Together, these solutions $\sigma$ are
the real ones that best approximate the Wilson fixed point. They are shown in
fig.\fig\sigmas{sigmas}\ up to $n=25$. (For the higher $n$ one must work to
an accuracy of at least 20 significant figures to avoid round off
errors). One clearly sees the four-fold
periodicity with amplitude $\approx\Delta$, and that, as expected,
 the upper solutions are slightly worse approximations.
Indeed the average of the last four solutions gives $\sigma=-.4607$
which is greater than $\sigma_*$ by $\approx\Delta/6$.
\midinsert
\centerline{
\psfig{figure=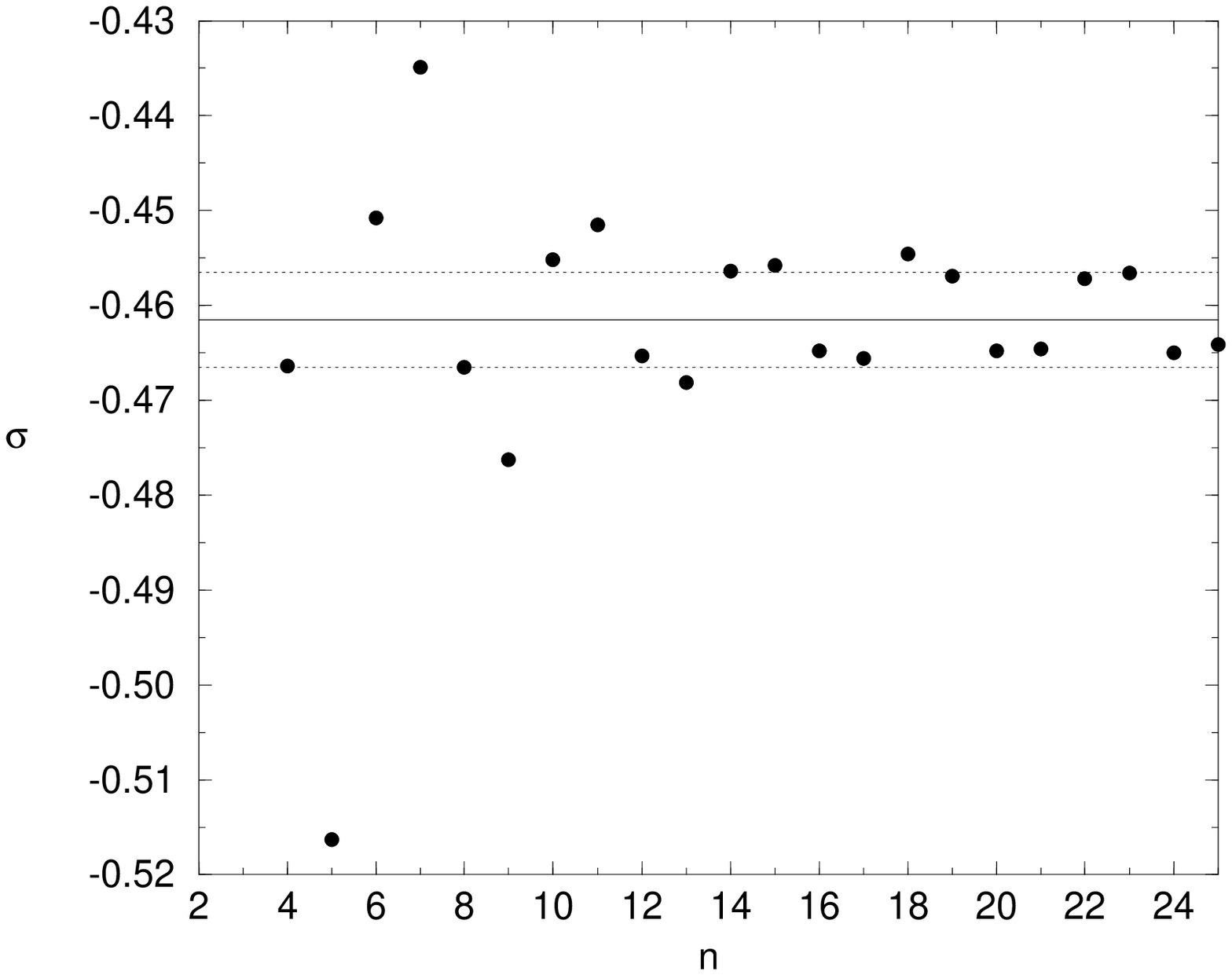,width=4.3in}}  
\vskip 0in
\centerline{\vbox{{\bf Fig.3.} The solutions $\sigma$ that best approximate
the exact answer ($\sigma=\sigma_*$, shown as a continuous horizontal line)
for the truncations $n$ up to $n=25$.
The solutions not displayed ($n=2,3$) lie outside the  range on the $\sigma$
axis.  The dotted lines are $\sigma=\sigma_*\pm.005$.
Recall that $\Delta\approx.005$.
}}
\endinsert

The four-fold
periodicity is transferred to the critical exponents, which may be
computed by linearizing, in \Veq, as
$\alpha_{2k}(t)=\alpha_{2k}(\sigma)+\epsilon\beta_{2k}\,\e{\lambda t}$
(where we have used separation of variables). The
neatest method of computing the eigenvalues $\lambda$ is by the
linear recurrence relations between the $\beta_{2k}$, imposing
$\beta_{2n+2}=0$. All truncations, but $n=23$, have one positive eigenvalue as
required, which yields\Wil\  $\nu=1/\lambda$.
The results are shown in fig.\fig\nus{nus}. One clearly sees again the
limiting periodic behaviour about the exact solution ($\nu=.6895$),
this time with
amplitude $\approx .008$. The exponent for the first correction to
scaling is given by $\omega=-\lambda$ where $\lambda$ is the least
negative eigenvalue. This depends much more sensitively on the
approximations and is even complex for $n=19,22,23$. It bounces
about the exact answer ($\omega=.5952$) with an amplitude $\approx .15$.
These truncations have been considered before, in
refs.\trunci\truncii\ to  $n=11,8$ respectively, but without the
deeper understanding
it was possible at these orders to interpret the numerical results as
indicating convergence. (The much worse behaviour for $O(N)$
scalar field theory with $N=2,3$ \truncii\ is surely due  to the exact solution
having complex singularities much closer to the origin).
\midinsert
\centerline{
\psfig{figure=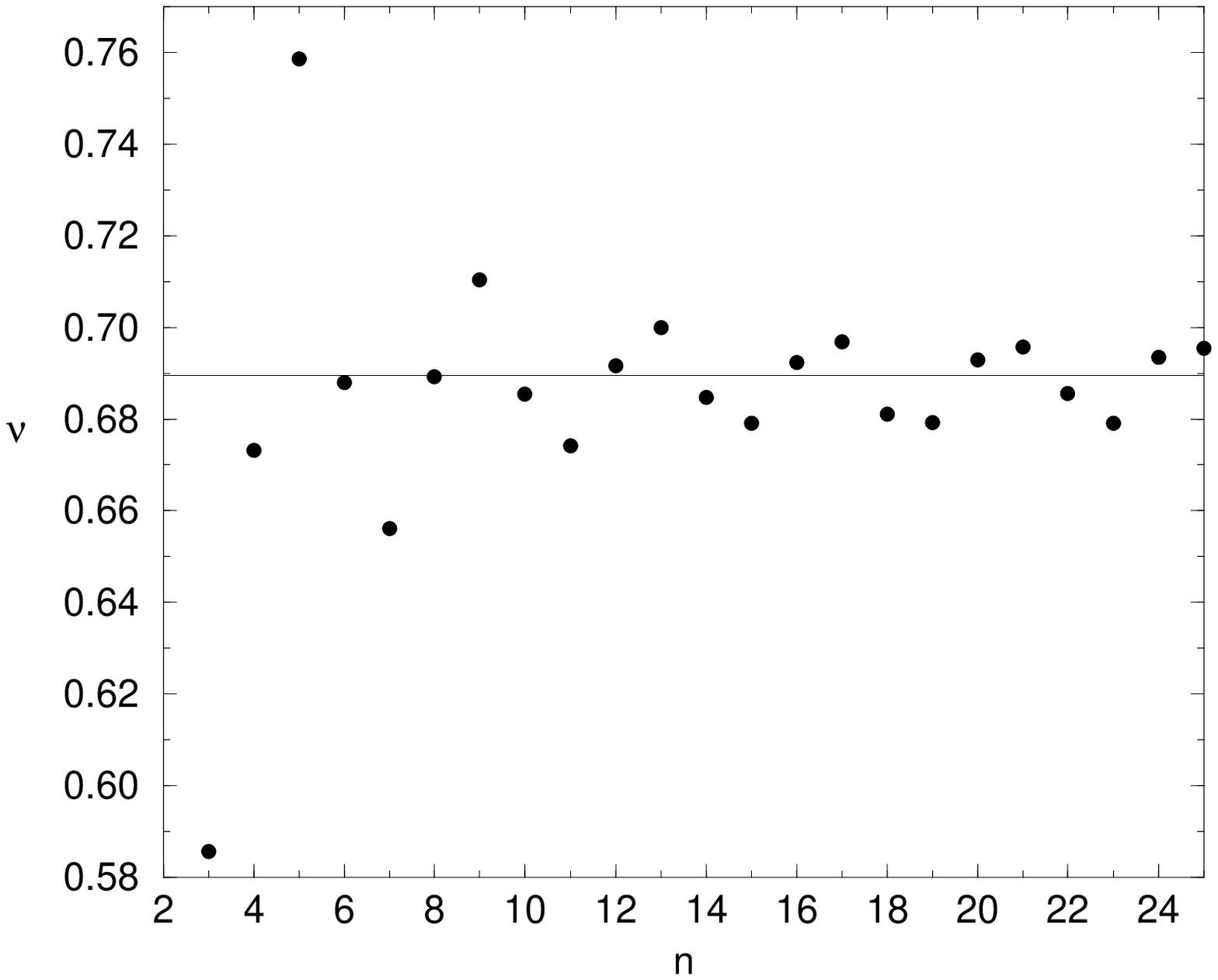,width=4.3in}}  
\vskip 0in
\centerline{\vbox{{\bf Fig.4.} The exponent $\nu$ for  truncations  up to
$n=25$. The exact answer is shown as a horizontal line.
}}
\endinsert

Of course these approximate solutions are not the only solutions for
$\sigma$. For $n$ too small the asymptotic pattern has not set in. Even
for $n$ large, there are many `spurious' solutions $\alpha_{2n+2}=0$
in the range $\sigma_*<\sigma<0$, which must be
there by continuity arguments;
 away from the boundaries of this range, one finds that they slowly
drift leftwards with increasing $n$ (asymptoting towards
$\sigma=\sigma_*$), reflecting the fact that the closest singularities
have an angle $\th$ which is a slowly
increasing function of $\sigma$ in this range.
Looking only at the truncations, how can one tell that
these solutions are spurious? There is surely no completely reliable method.
 There is no good reason
to require the {\sl truncated} potential to be bounded below and
indeed the first cases to violate that criterion are truncations $n=6,7$
in figs.\sigmas,\nus; nor is it compelling to assume the
solutions must be real: the approximations are bad (compared to their
neighbours) for truncations 22 and 23, unless one chooses
certain solutions with small
imaginary parts ($\sigma=-.4572+.0059i$ and $-.4566+.0027i$ respectively --
only the real parts are shown in figs.\sigmas,\nus). For
$n=23$  even the requirement that the approximation should have
only one relevant direction, breaks down: there are no such solutions.
The case we chose has $\lambda=1.472+.0253i$, which we use to
compute $\nu=.6791-.0117i$, and a less relevant direction with
$\lambda=.509+1.695i$. The spurious solutions generally, {\sl but not always},
have about the same number of positive eigenvalues $\lambda$ as negative
eigenvalues: which is many for large $n$.
The best one can do\trunci\ to eliminate spurious solutions,
is to look numerically for convergence/stability with
increasing $n$. That this is not good
enough is nicely demonstrated in ref.\trunci\ where the slow drift of a
sequence of spurious solutions $\sigma$, with two relevant eigenvalues,
is mistaken for convergence to a tricritical point.

Finally we mention two better expansion methods. The first is the
analytic equivalent of shooting to an intermediate fitting point, namely
we require the Taylor expansion \expa\ and its derivative to agree
with the asymptotic expansion \asy\ and its derivative, at some given
intermediate point $\phi_f$. In this way one obtains just one
solution as expected, with $\sigma\approx\sigma_*$ and $A\approx A_*$.
 By comparing a pair of truncated Taylor expansions
with a pair of truncated asymptotic expansions over a range of $\phi_f$
(to, at the same time, estimate the error and determine the `best' $\phi_f$)
one can extract moderately accurate bounding values for $\sigma_*$,
for example within a range of width .04 by using Taylor series to $\phi^8$
and $\phi^{10}$ and asymptotic series to $\phi^{-10}$ and $\phi^{-14}$.
This method will be
reliable providing the asymptotic series is accurate for $\phi>a$
(when $A\approx A_*$) for some $a$ such that $a<r_*$,
which is certainly case here.
Presumably it works in general if there are no singularities in the
region $a<{\rm Re}(\phi)<\infty$, except that it cannot give unlimited
accuracy since the expansion \asy\ converges only in the Poincar\'e sense.
\midinsert
\centerline{
\psfig{figure=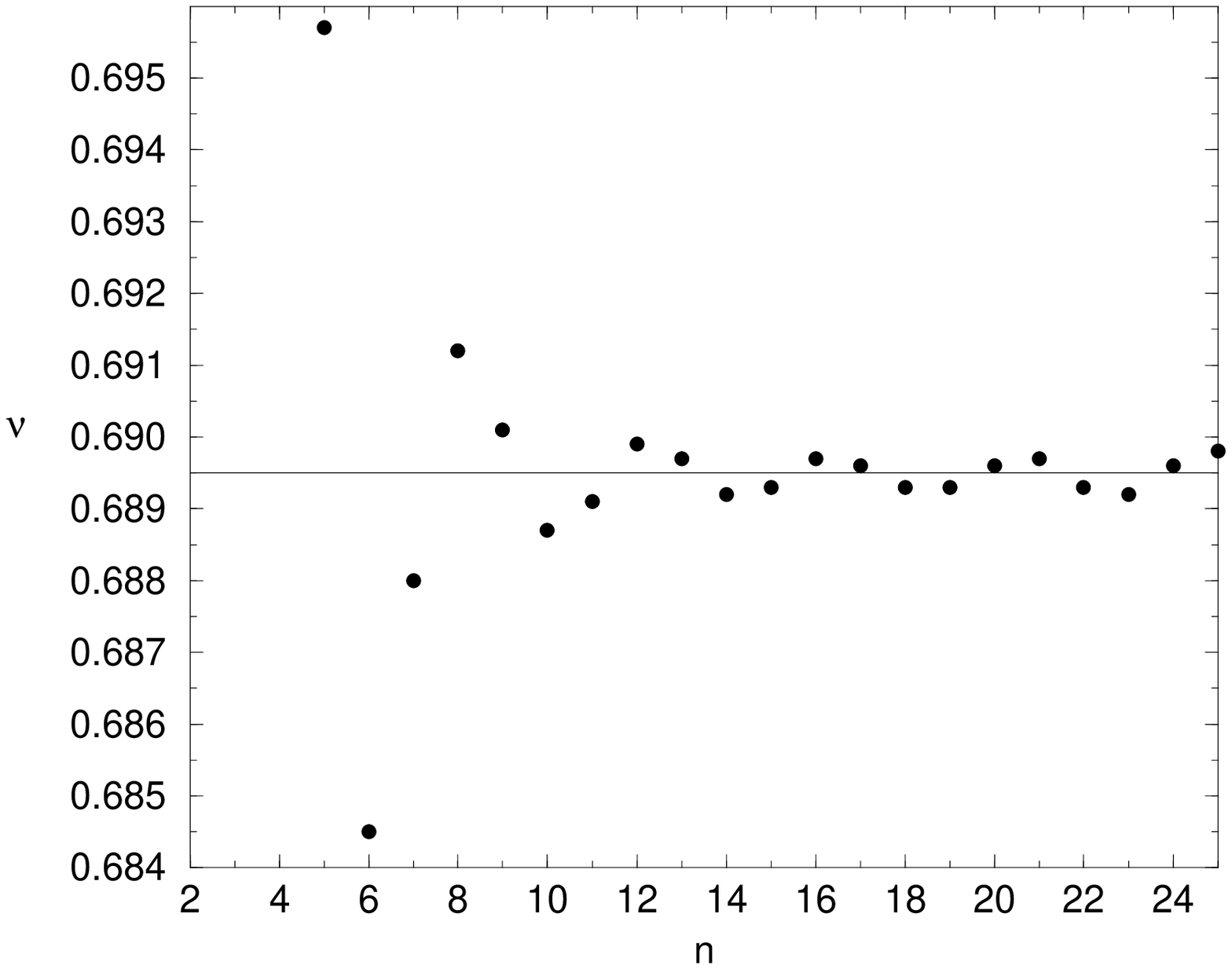,width=4.3in}}  
\vskip 0in
\centerline{\vbox{{\bf Fig.5.} Results for the exponent $\nu$ against $n$,
in an expansion method that utilises the fact that $\th_*\approx\pi/4$.
 The exact answer is shown as a horizontal line. Note the much finer vertical
scale compared to fig.4.
}}
\endinsert

In the second method we assume that $\sigma_*$ has been determined to the
accuracy required of the eigenvalues.
Linearising about the fixed point position of the
nearest singularity as $\th=\th_*+\epsilon \psi\,\e{\lambda t}$
and $r=r_*+\epsilon s\,\e{\lambda t}$, one obtains the asymptotic
behaviour of $\beta_{2k}$:
$$\beta_{2k}\sim -4k\left\{s\cos(2k\th_*)/r_*^{2k+1}
+\psi\sin(2k\th_*)/r_*^{2k}\right\}\quad .$$
Thus to leading order in $1/n$, 
$$(n-1)\beta_{2n+2}\alpha_{2n-2}(\sigma_*)-(n+1)\beta_{2n-2}
\alpha_{2n+2}(\sigma_*)\sim -8 (n^2-1)\psi \sin(4\th_*)/r_*^{4n}\ .$$ 
Since $\th_*\approx\pi/4$ we can expect that it is a good approximation
to choose $\lambda$ such that the left hand side vanishes. Doing so, we
find a spectacular improvement in convergence, cf. fig.\fig\nushift{  },
which nicely provides further
confirmation for our theory.  By looking at running averages of four points,
and assuming no  systematic shift from neglecting the right hand side of the
above, we extract $\nu=.689457(8)$. Similarly we find $\omega=.5955(5)$.

\acknowledgements

It is a pleasure to thank the following people for their interest and
discussions:
Richard Ball, Michel Bauer, Simon Catterall, Poul Damgaard, Patrick Dorey,
Dan Freedman, Peter Haagensen, Tim Hollowood, Yuri Kubyshin,
Jose Latorre and Ulli Wolff.

\listrefs
\end